\documentclass[conference]{IEEEtran}
\IEEEoverridecommandlockouts
\usepackage{cite}
\usepackage{amsmath,amssymb,amsfonts}
\usepackage{graphicx}
\usepackage{textcomp}
\usepackage{xcolor}

\usepackage{algorithm}
\usepackage[noend]{algpseudocode}
\usepackage{tcolorbox}
\tcbuselibrary{skins,breakable}
\algrenewcommand\algorithmicrequire{\textbf{Input:}}
\algrenewcommand\algorithmicensure{\textbf{Output:}}

\DeclareMathOperator*{\argmin}{arg\,min}
\DeclareMathOperator*{\E}{\mathbb{E}}
\newcounter{algcounter} 

\newtcolorbox{algobox}[2][]{%
  enhanced,
  breakable,
  colback=white,
  colframe=black,
  fonttitle=\bfseries,
  title={Algorithm~\refstepcounter{algcounter}\thesubsection: #2},
}

\def\BibTeX{{\rm B\kern-.05em{\sc i\kern-.025em b}\kern-.08em
    T\kern-.1667em\lower.7ex\hbox{E}\kern-.125emX}}
\begin{document}

\title{Robust Rule-Based Sizing and Control of Batteries for Peak Shaving Applications
\thanks{This research was funded in whole by the Swiss National Science Foundation (SNSF) 10DU--$\_$224166 in the context of the Grid-Aware Decarbonization of electricity-driven Neighbourhoods (GARDEN) project. For the purpose of open access, a CC BY public copyright licence is applied to any author accepted manuscript (AAM) version arising from this submission.}
}

\author{\IEEEauthorblockN{Lorenzo Nespoli}
\IEEEauthorblockA{\textit{ISAAC} \\
\textit{SUPSI}\\
Mendrisio, Switzerland \\
lorenzo.nespoli@supsi.ch}
\and
\IEEEauthorblockN{Vasco Medici}
\IEEEauthorblockA{\textit{ISAAC} \\
\textit{SUPSI}\\
Mendrisio, Switzerland \\}
}

\maketitle

\begin{abstract}
As the cost of batteries lowers, sizing and control methods that are both fast and can achieve their promised performances when deployed are becoming more important. In this paper, we show how stochastically tuned rule based controllers (RBCs) can be effectively used to achieve both these goals, providing more realistic estimates in terms of achievable levelised cost of energy (LCOE), and better performances while in operation when compared to deterministic model predictive control (MPC). We test the proposed methodology on yearly profiles from real meters for peak shaving applications and provide strong evidence about these claims.  
\end{abstract}


\section{Introduction}

\subsection{Related works on RBCs for battery management}
Previously proposed RBCs for electrical batteries have been enthusiastically received by the academic and industrial communities. The success is grounded in the fact that specific problems in energy storage management can be proven to have a simple solution structure. For example, in \cite[Theorem 2]{chen_electric_2014} authors show that any EV charging scheduling problem with a convex objective function and requiring a unitary state of charge (SoC) at the end of the optimization horizon has an optimal solution with a valley-filling structure. In \cite[Theorem 2]{shi_optimal_2019} the authors prove near optimality for an online algorithm with a two threshold structure for the revenue maximization in frequency regulation markets of an electric battery, under a convex objective function and convex degradation models. In \cite[Theorem 1]{van_de_ven_optimal_2013} authors show that optimal battery management under the arbitrage problem has a two-threshold structure. In all the aforementioned cases, the authors didn't provide closed form solutions for the thresholds, which they find numerically using dual decomposition of the power dispatch problem, stochastic gradient descent, and policy iteration, respectively. Stochastic tuning of RBCs has been proposed, for example, in \cite{ghadimi_stochastic_2024}, where the authors proposed a stochastic gradient method for this task.

\subsection{Background on policy optimization}

Storage controllers are defined as policies $\pi$ mapping the current set of available information $z_t = (p_t, x_t) \in  \mathbb{R}^{d+1}$, where $p_t\in \mathbb{R}$ is the current uncontrolled power and $x_t \in \mathbb{R}^d$ a vector of additional features, to control actions in terms of battery power, $p_{b, t} = \pi(z_t)$. Some policies are generated via (stochastic) online optimization with look-ahead, like MPC or tree-based stochastic MPC (TB-SMPC). These look-ahead policies usually rely on a forecast of $p$, $\hat p(z)$, to plan over a limited time horizon. Other techniques, e.g.  reinforcement learning, obtain policies as solutions of a stochastic optimization over a static, usually much bigger, training set. We can describe these three mentioned policies and stochastically tuned RBC as per equation \eqref{eq:controllers}: 

\begin{equation}\label{eq:controllers}
p_{b,t_k}^* = \begin{cases}
p_b^*(0) \vert \ p_{b}^* \in \argmin\limits_{p_b \in \mathcal{P}} \sum\limits_{t\in \mathcal{T}_h}l_h(\hat{p}_t(z_t), p_{b, t}) \ & \text{MPC}\\
p_b^*(0) \vert \ p_{b}^* \in\argmin\limits_{p_b \in \mathcal{P}} \sum\limits_{n\in \mathcal{T}_{h,n}}l_h(\hat{p}_{n}(z_n), p_{b,n})\ &    \text{TB-SMPC}\\
\pi_{\theta^*}(z_{t_k})\vert \ \theta^* \in  \argmin\limits_{\pi_\theta(z)\in \mathcal{P}} \E\limits_{\ \mathcal{D}_{tr}} l_p(p_t, \pi_\theta(z_t)) \ \   &\text{RL}, \text{RBC}
\end{cases}
\end{equation}

where $\mathcal{T}_h$ is the set of time-steps over a limited optimization horizon starting at $t_k$ over which the loss $l_h$ is computed, usually covering one or two days for this kind of applications, $\mathcal{T}_{h, n}$ is a set of nodes indicating a tree-based representation of the forecasted power profile, with a root node at $t_k$ and $\mathcal{P}$ is a set of operational constraints. The set $\mathcal{D}_{tr} = \{z_t\}_{t \in \mathcal{T}_{tr}}$ indicates the dataset of powers and possible features over a much longer period, e.g. 6 months, over which the loss $l_p$ is computed. As we can see from equation \eqref{eq:controllers}, optimizing RBCs stochastically is conceptually similar to RL. The main difference is the functional form used to constrain the search for optimal policies. While for (non-tabular) RL $\pi_\theta$ is usually a neural network, for rule based controllers the functional form $\pi_\theta$ is usually much simpler and encodes some domain knowledge of the control problem. 

\subsection{Contributions}
This paper makes three contributions.
\begin{itemize}
    \item In section \ref{sec:RBC}, we propose a simple parametric RBC for peak shaving whose parameters can be efficiently tuned via gradient-free optimization. Since consumption profiles are usually not stationary, we propose to find decision thresholds using simple parametrized running statistics.
    \item In section \ref{sec:RRBC}, we introduce a robust training objective based on Conditional Value-at-Risk (CVaR) applied to daily peaks, improving the generalization of RBCs under non-stationary consumption.
    \item Finally, in section \ref{sec:results}, we show that using these RBCs inside the sizing problem produces more realistic LCOE estimates and smaller performance gaps between sizing-time and ex-post evaluations, outperforming deterministic MPC under realistic forecasting errors.
\end{itemize}

The proposed RBCs are intentionally simple, interpretable, and inexpensive to tune, which facilitates their deployment in industrial settings.

\section{Methodology}\label{sec:methodology}
\subsection{Problem formulation}
In this paper, we consider a tariff scheme with a flat buying price and a monthly peak tariff. A usual objective when sizing a battery is to reduce the levelised cost of energy (LCOE), which could be expressed as a function of capex ($\mathrm{C}$) and opex ($\mathrm{O}$): 
\begin{align}
\mathrm{C}&=c_{\mathrm{bat}}^E E_{\mathrm{bat}}+c_{\mathrm{bat}}^P P_{\mathrm{bat}}\label{eq:capex}\\
\mathrm{O}&=\Delta t \sum_{t\in\mathcal{T}_{tr}}\left(\lambda_t^{\mathrm{imp}} p_t^{+} -\lambda_t^{\exp} p_t^{- } \right) + \sum_{m \in \mathcal{M}} \lambda^{\text{p}} p_m^{\text {peak }} \label{eq:opex}\\
\mathrm{LCOE} &= \frac{\operatorname{crf} \mathrm{C} + \rho \mathrm{O}}{\rho E}\label{eq:lcoe}
\end{align}
The capex is computed from the price of the battery pack times its size and the inverter's price times its size (terms appearing in this order in equation \eqref{eq:capex}). The opex $\mathrm{O}$ can be computed knowing the operations of the battery $p_{b,t}$, affecting the imported power $p_t^{+} = \max{(p_t + p_{b,t}, 0)}$, the exported power $p_t^{-} = -\min{(p_t + p_{b,t}, 0)}$ and the monthly peak $p_m^{peak} = \max_{t\in\mathcal{T}_m} p_t+p_{b,t}$, $\mathcal{T}_m$ being the set of time-steps belonging to month $m$ and $\mathcal{M}$  being the set of months in the training set, weighted for their respective prices $\lambda_t^{\mathrm{imp}}, \lambda_t^{\mathrm{exp}}, \lambda^{p}$. In equation \eqref{eq:lcoe} the $\mathrm{crf}$ is the actualization factor $\mathrm{crf} = \frac{r(1 + r)^n}{(1 + r)^n - 1}$ where $r$ is the interest rate and $n$ is the expected lifetime of the investment. Since the opex $O$ is usually estimated using less than one year of operations by selecting just $\tau$ typical days, $\rho = \frac{365}{\tau}$ is a multiplicative factor for estimating the yearly costs and yearly consumed energy from the partial estimation $E=\sum_{t\in\mathcal{T}_{tr}}p_t\Delta t$. 

The sizing problem can then be written as a joint optimization over the battery capacity, power and battery operations $p_b$:
\begin{align}
\argmin_{p_{b},E_{\mathrm{bat}} P_{\mathrm{bat}}} & \quad \mathrm{LCOE}(p, p_b, E_{\mathrm{bat}},P_{\mathrm{bat}} ) \label{eq:sizing}\\
s.t.& \quad e_{t+1} = e_{t} + \eta_{ch}p_{b}^{ch} - \frac{1}{\eta_{ds}}p_{b}^{ds}\\
& 0 \leq p_b^{ch} \leq P_{\max}, 0 \leq p_b^{ds} \leq P_{\max} \\
&E_{min} \leq e \leq E_{max}
\end{align}

where $p_b^{ch} = \max(p_b, 0)$ and $p_b^{ds} = \max(-p_b, 0)$
are the charging and discharging powers and $P_{max}, E_{min}, E_{max}$ are power and energy operational limits respectively. 
Problem \eqref{eq:sizing} can be solved in one shot, retrieving perfect operations ad size, if we assume to perfectly know the future in terms of the power at meter, $p$. The alternative would be to perform a nested optimization in which a master problem chooses the system sizing ($E_{\mathrm{bat}} P_{\mathrm{bat}}$), while the inner problem solves the operational problem using e.g. realistic forecasts.
This can be done, but it requires the interplay of a forecaster and a (possibly multistage stochastic) controller, which makes the overall optimization expensive and subject to modeling errors. Furthermore, optimizing for the monthly peaks requires a long planning horizon or to keep track of the peak so far during the current month. In the following section, we explain how this joint optimization becomes much more attractive for parametric rule-based controllers.

\subsection{Stochastically tuned RBCs for peak shaving}\label{sec:RBC}
We propose a simple RBC with three parameters, $\theta \in \mathbb{R}^3$, whose pseudo-code is reported in Algorithm~\ref{alg:rbc}. Defined $z_t = p_{t\vert t-\theta_1} \in \mathbb{R}^{\theta_1}$ as the collection of the last $\theta_1$ unconstrolled powers, the battery discharges if $p_t$ is above the $\theta_2$-quantile estimation on $z_t$, $q_t^u = q_{\theta_2}[z_t]$. It does so respecting the battery operational limits and such that the overall power doesn't fall below $q_t^u$. Discharging happens with similar constraints when $p_t$ falls below the similarly defined $q_t^l$.

\begin{algobox}{RBC for peak shaving}
\label{alg:rbc}
\begin{algorithmic}[1]
\algrenewcommand\alglinenumber[1]{\scriptsize #1} 
\Require $\theta$, $z_t = p_{t\vert t-\theta_1} \in \mathbb{R}^{\theta_1}$
\Ensure $\pi_t = p_{b,t}$

\State $q^u_t \gets q_{\theta_2}[z_t]$, $q^l_t \gets q_{\theta_3}[z_t]$
    \If{$p > q^u_t$}    \Comment{discharge}
        \State $p_{b,t} \gets \min\{\,p-q^u_t,\;P_{\max},\;\frac{E-E_{\min}}{\eta_{dis}/\Delta t}\,\}$

    \ElsIf{$p < q^l_t$} \Comment{charge}
        \State $p_{b,t} \gets \min\{\,q^l_t - p_t,\;P_{\max},\;\frac{E_{\max}-E}{\eta_{ch}\Delta t}\,\}$
    \Else               \Comment{idle}
        \State $p_{b,t} \gets 0$
    \EndIf
\end{algorithmic}
\end{algobox}
Simulating Algorithm~\ref{alg:rbc} is fast. When compiled with $\texttt{numba}$\footnote{https://numba.pydata.org/} this RBC takes around $120 \pm 9 \mu s $ to simulate one year at hourly resolution. As a comparison, the MPC optimization we use in section \ref{sec:results} needs $44s$ for the same period of time. This makes it convenient to tune $\theta$ stochastically on $\mathcal{D}_{tr}$ using a gradient-free solver.

\subsection{Robust RBCs via Conditional Value ar Risk}\label{sec:RRBC}
We increase the generalization of the RBC by crafting a surrogate daily loss $l_d$ and minimizing it robustly. Instead of tuning the RBC to shave the monthly peaks or minimize the LCOE, we can fit its parameters to minimize the right tail of the distribution of $l_d$ using an archive of losses $\mathcal{D}_{tr}^d = \{l_d\}_{d=1}^D$ built using the training set $\mathcal{D}_{tr}$. The easiest daily loss in this case would be the daily maximum after battery operations:
\begin{equation}\label{eq:daily_peak}
l_d = \max_{t\in\mathcal{T}_d} p_t+p_{b,t}
\end{equation}
where $\mathcal{T}_d$ is the set of time-steps belonging to day $d$.    
We can minimize the tail risk using as an objective the conditional (or tail) value at risk, CVaR \cite{artzner_coherent_1999}:
\begin{align}
\operatorname{CVaR}_\alpha(L)&=\min _{y \in \mathbb{R}}\left\{y+\frac{1}{1-\alpha} \mathbb{E}\left[(L-y)^{+}\right]\right\} \label{eq:cvar_min}\\
&\simeq\sup _{w \in \mathcal{W}_a} \sum_{d=1}^D w_d \ell_d \label{eq:cvar_dual}\\
&=\frac{1}{k} \sum_{d=1}^k \ell_{(d)} \label{eq:cvar_layman}
\end{align}

where $L$ is the random variable associated to $l_d$,  $\mathcal{W}_\alpha=\left\{w \geq 0, \sum_d w_d=1, w_d \leq \frac{1}{(1-\alpha) D}\right\}
$, $k = (1-\alpha)D$ where $D=\vert \mathcal{D}_{tr}^d\vert$ is the cardinality of the daily training dataset and $\cdot_{(d)}$ is the descending order permutation $l_{(1)} \geq l_{(2)} \geq \cdots \geq l_{(D)}$. We briefly explain the above equations. It is well known that, if there are no atoms at $F^{-1}_\alpha(L)$, equation \eqref{eq:cvar_min} is equal to $\operatorname{CVaR}_\alpha(L)=\mathbb{E}\left[L \mid L \geq F^{-1}_\alpha(L)\right]$, which justifies the minimization of  $\operatorname{CVaR}_\alpha(L)$ to control tail risk. Equation \eqref{eq:cvar_dual} is the empirical finite sample estimator of the dual form introduced in \cite{rockafellar_optimization_2000}. We can interpret it as an adversarial re-weight of the empirical day peak distribution, with capped weight at $\frac{1}{(1-\alpha) D}$, $\mathcal{W}_\alpha$ being the empirical risk-envelope \cite[Proposition 6.4]{royset_risk-adaptive_2025}. In fact, minimizing \eqref{eq:cvar_dual} w.r.t. the policy parameters $\theta$ is equivalent to solve a distributionally robust optimization \cite{royset_risk-adaptive_2025}. Finally, \eqref{eq:cvar_dual} boils down to the more practical expression \eqref{eq:cvar_layman}, stating  $\operatorname{CVaR}_\alpha(L)$ can be estimated by averaging the $(1-\alpha) D$ more extreme peaks.  
Since consumption profiles can be seasonal, the most challenging peaks could be located in the same (usually winter) month. To avoid seasonal dominance, we used a monthly stratified CVaR:
\begin{equation}\label{eq:scvar}
    \mathrm{SCVaR}_\alpha=\frac{1}{M} \sum_{m=1}^M \frac{1}{k_m} \sum_{i=1}^{k_m} \ell_{(i)}^{(m)}
\end{equation}
where $M = \vert \mathcal{M}\vert$ and $k_m=\left\lfloor(1-\alpha) D/M\right\rfloor$.

\subsection{Tariff and Financial Assumptions}

For the customer size considered in this work, we assume a tariff that reflects the general trend towards more peak-oriented grid charges. As baseline, we take a flat import price of 200~USD/MWh, of which 70~USD/MWh are attributed to grid usage, and a monthly peak charge of 5750~USD/MW/month on the maximum hourly average power in each calendar month. To emulate a future, more peak-oriented but revenue-neutral grid tariff, we shift one-half of the grid-related volumetric charge from energy to the monthly peak component and calibrate this shift on the training data so that the total annual grid revenue remains unchanged. This results in an effective import price of 165~USD/MWh and a monthly peak tariff of 20\,044~USD/MW/month, which are used in all sizing and control experiments when computing the opex term in \eqref{eq:opex}.

Battery investment costs are modelled at 120~USD/kWh for energy capacity, 50~USD/kW for power capacity, and a fixed project cost of 1\,000~USD per installation. We assume a project lifetime of 15~years and a discount rate of 6\,\%, which defines the capital recovery factor in the LCOE calculation.

\section{Numerical results}\label{sec:results}
To test our approach, we used a dataset of Portuguese medium voltage energy meters \cite{trindade_electricityloaddiagrams20112014_2015}. The original dataset included 370 series spanning 3 years. We downsampled the original 15-minute resolution to 1 hour and restricted the analysis to the first 100 power meter series, and used one year of data to speed up computation.
For each series, we sized the battery using two sizing methods on a training set $\mathcal{D}_{tr}$ spanning the first 6 months of the year, and tested their promises in terms of LCOE on a test dataset $\mathcal{D}_{te}$ using the second part of the year. 
The sizing methods are the following:
\begin{enumerate}
    \item {\bf{Prescient}} sizing. This sizing method assumes that the battery controller perfectly knows the future. This makes the sizing problem \eqref{eq:sizing} solvable in a reasonable amount of time but makes the final sizing optimistic in terms of performances. We solved this problem via \texttt{GUROBI}\footnote{https://www.gurobi.com/}.
    \item {\bf{RBC}} sizing. We optimize \eqref{eq:sizing} using the RBC of Algorithm~\ref{alg:rbc}. To this end we used the gradient-free \texttt{differential$\_$evolution} solver from scipy\footnote{https://scipy.org/}. This process also returns an optimal set of control parameters  $\theta^*_{\mathrm{sizing}}$ 
\end{enumerate}
The tested controllers are described in the following:
\begin{enumerate}
    \item MPC. We solve the corresponding problem in \eqref{eq:controllers} in a receding horizon fashion where $l_h = (\hat p_t(z_t)+p_{p, t})^2$ is a quadratic peak shaving objective of the total profile over a 24-hour horizon. To ensure fairness, MPC uses the same battery model and operational constraints as the RBC, including identical limits, efficiencies, and SoC bounds. The predictions $\hat p_t(z_t)$ were obtained by training 24 different LightGBM models, where $z_t$ contains the previous 24 hours' lags for $p_t$ and calendar features. This kind of modeling has been proven effective for aggregated demand profile forecasting \cite{nespoli_hierarchical_2020}. The median and 90\% quantile normalized MAE across meters on test set are 0.102 and 0.193, respectively. Each optimization horizon is solved via \texttt{GUROBI} using warm start.
    \item Prescient MPC. It solves the same problem of the previous controller, but makes perfect forecasts over the horizon. As such, it can be considered a lower bound for daily peak shaving and a very good solution when optimizing for monthly peak tariffs.
    \item RBC. The RBC, whose parameters $\theta$ are either optimized stochastically on $\mathcal{D}_{tr}$ to minimize the average daily peak from equation \eqref{eq:daily_peak}, when the sizing is done using the prescient method, or are set to $\theta^*_{\mathrm{sizing}}$ when the RBC sizing is done instead.
    \item Adversarial RBC. In this case $\theta^*$ is obtained minimizing \eqref{eq:scvar} with level $\alpha=0.95$.   
\end{enumerate}
For the last two options we used differential evolution as optimizer, as we did for the RBC sizing.
\begin{figure}[h]
\centering
\includegraphics[width=1\linewidth]{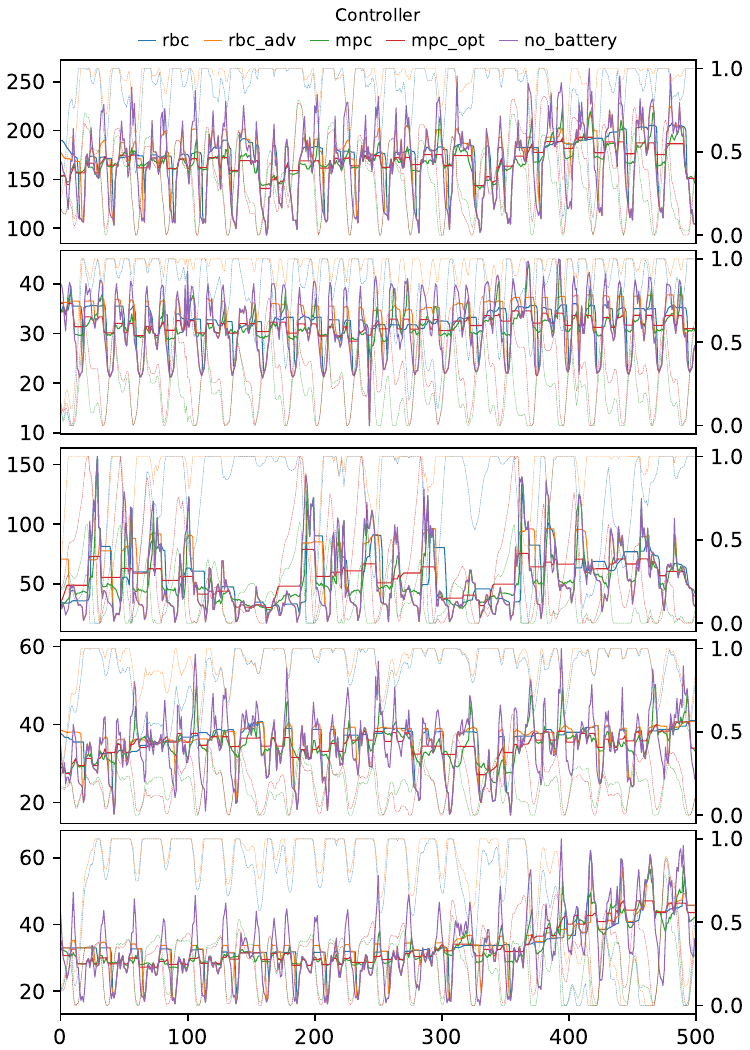}
\caption{Uncontrolled (violet) and controlled profiles for 6 time series and four compared controllers. Dashed lines represent the SoC of the battery. Left axis: power, kWh; right axis SoC.}
\label{fig:opt_examples}
\end{figure}

Figure \ref{fig:opt_examples} shows examples on 5 time series in terms of optimized profile $p_t + p_{b,t}$ and SoC for the four controllers, over 500 hours. It can be noticed that the two RBCs tend to keep the battery fully charged for long periods of time, which could be sub-optimal. The adversarially optimized RBC always chose a higher quantile for the computation of the upper threshold $q_{\theta_2}$. This makes the policy more conservative and, in general, more robust to extreme peaks. The last days of the fifth row show cases in which the RBCs are particularly challenged: if consumption profiles are non stationary, the running quantile heuristic doesn't excel during transition periods.

\subsection{Effect on peaks}
\begin{figure}[h]
\centering
\includegraphics[width=1\linewidth]{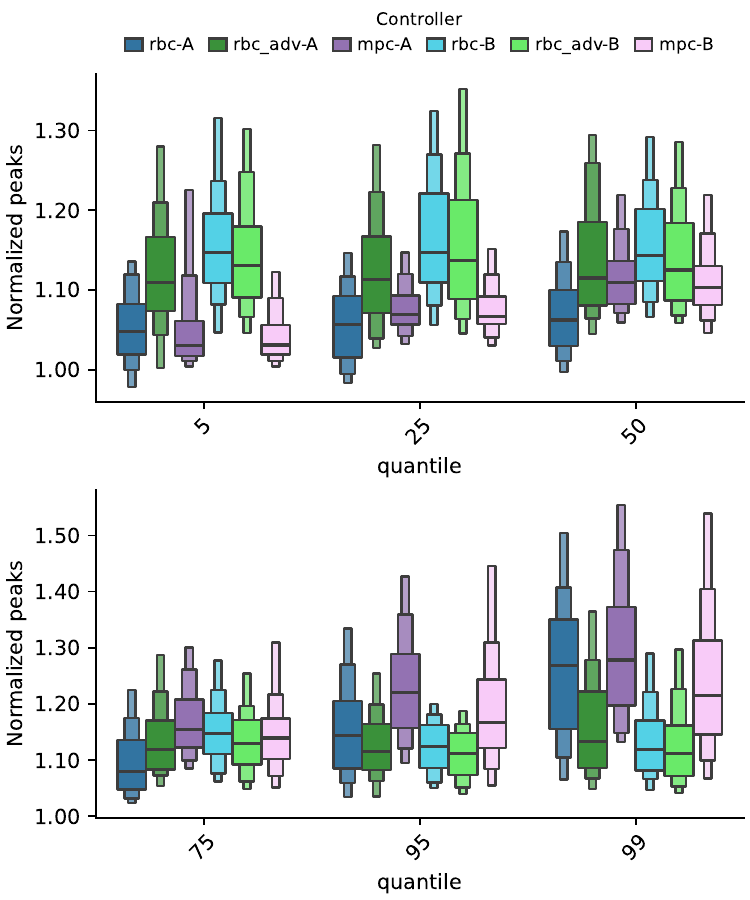}
\caption{Boxen plots of quantiles of normalized daily maxima across meters, for different controllers and sizing methods. A: prescient sizing. B: RBC sizing. Top: lower quantiles. Bottom: higher quantiles. Outliers are not shown.}
\label{fig:quantiles}
\end{figure}
To investigate robustness for peak shaving applications, we assess the performance in shaving the daily peaks. We define the normalized daily peak as $\frac{l_d(\pi)}{l_d(\pi_{\mathrm{MPC-pre}})}$ where $l_d(\pi)$ is the daily peak loss defined in \eqref{eq:daily_peak} and $\pi_{\mathrm{MPC-pre}}$ is the prescient MPC. Figure \ref{fig:quantiles} shows the normalized daily peak distributions over the 100 series for increasing quantiles, for different controller and sizing methods. Dark colors refer to the prescient sizing, while lighter to RBC sizing. While MPC achieves lower distributions w.r.t. the RBC controllers in the lower quantiles, indicating a superior ability in lowering smaller peaks, for higher quantiles, RBCs are, on average, more effective. When considering the 95th and 99th quantiles, the adversarially tuned RBC is the best controller. However, under RBC sizing, the performances of the two RBCs become similar. 

\subsection{Effect on sizing}
\begin{figure}[h]
\centering
\includegraphics[width=0.8\linewidth]{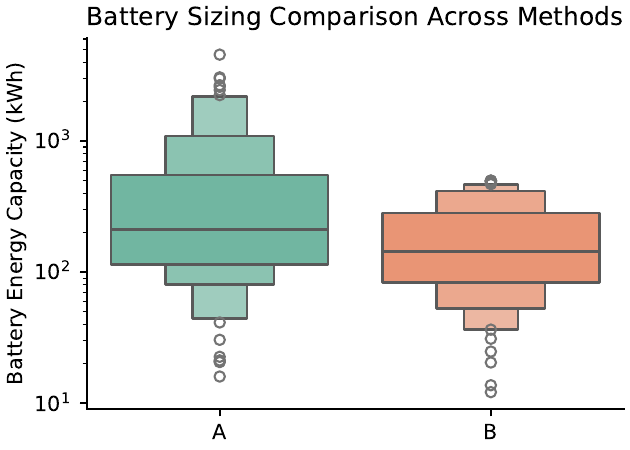}
\caption{Battery size boxen plots across series with the two sizing strategies. Left: prescient sizing; right: RBC sizing. Log scale. }
\label{fig:sizing}
\end{figure}

Since the prescient optimization is optimistic in terms of battery performances, it tends to oversize the battery. In comparison, the RBC is more conservative. Figure \ref{fig:sizing} shows the distributions for the battery sizes across meters, for both the sizing strategies. 

\subsection{Effect on LCOE}
\begin{figure}[h]
\centering
\includegraphics[width=1\linewidth]{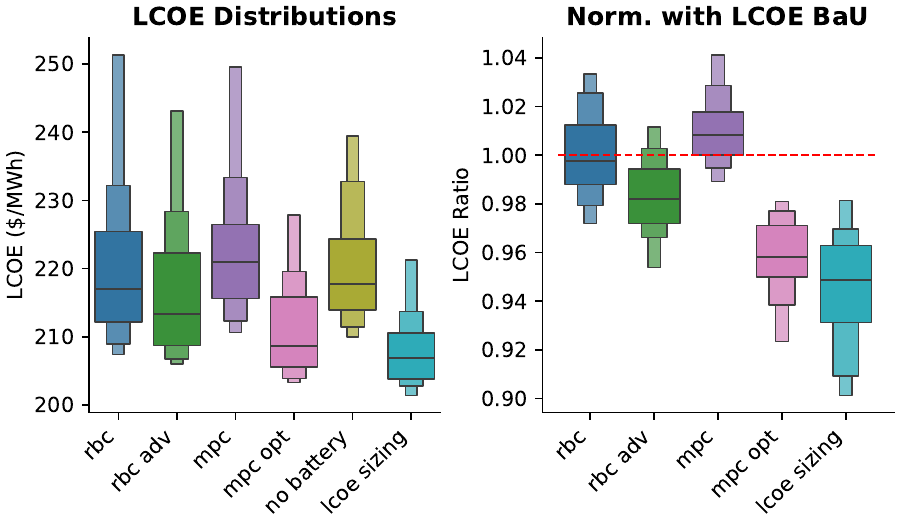}
\caption{Boxen plots across meters of LCOE on the test set, under prescient sizing. Left: unnormalized distribution; the distribution of the LCOE estimated on the sizing set is shown in violet. Right: LCOE distributions normalized with the distribution of the LCOE estimated on the sizing set.}
\label{fig:lcoe_prescient}
\end{figure}

\begin{figure}[h]
\centering
\includegraphics[width=1\linewidth]{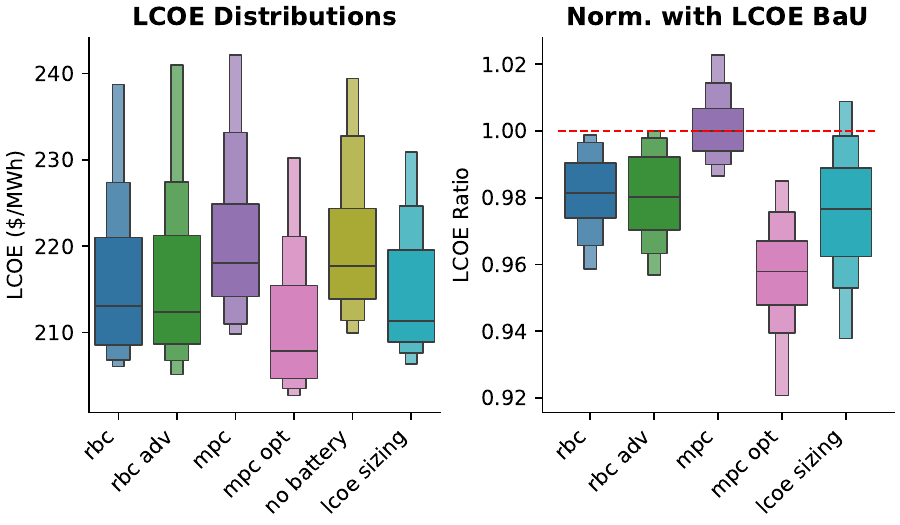}
\caption{Boxen plots across meters of LCOE on the test set, under RBC sizing. Left: unnormalized distribution; the distribution of the LCOE estimated on the sizing set is shown in violet. Right: LCOE distributions normalized with the distribution of the LCOE estimated on the sizing set.}
\label{fig:lcoe_rbc}
\end{figure}

Figure \ref{fig:lcoe_prescient} shows the LCOE distributions across meters for different controllers when using the prescient sizing method. All the boxenplots refer to the post-sizing performances on the test set, but the last one (cyan) refers to the LCOE evaluated at sizing time (in this case for the prescient sizing). It can be seen that, in general, LCOE is underestimated at sizing time. This once again can be explained by the prescient sizing method's optimism. The right panel of Figure \ref{fig:lcoe_prescient} shows the same distributions but normalized by the LCOE estimated on the training set under business as usual (BaU), that is, without a battery. Values greater than 1 thus indicate the installation of the battery not being profitable using an ex-post, counterfactual evaluation. It can be seen that while almost half of the cases for the RBC and the majority of them for the MPC are disattending promises in terms of LCOE, the adversarially tuned RBC provides lower LCOEs for most of the cases.  
Figure \ref{fig:lcoe_rbc} shows the same plots for the RBC sizing case: both RBC controllers successfully decrease LCOE w.r.t. BaU, while MPC fails in doing so for half the meters.

\section{Conclusions}
In this work we investigate the use of stochastically tuned parametric RBCs as a way to take into account stochasticity at sizing time, as opposed to deterministic, prescient sizing. We show that this can be done efficiently by using gradient-free optimization. By better representing operational performances, RBC sizing opts for smaller system sizes, but lowering the distances between predicted LCOE at sizing time and ex-post estimations. This shows RBCs can be used to better estimate the return on investment for this kind of applications, while being simple, interpretable, and computationally inexpensive to tune and deploy.

\bibliographystyle{IEEEtran}
\bibliography{references}
\end{document}